\documentclass{aa}  

\usepackage{graphicx}
\usepackage{txfonts}
\usepackage[draft]{hyperref}
\def \lsun{\ifmmode{{\rm\ L}_\odot}\else{${\rm\ L}_\odot $}\fi}

\def \msun{\ifmmode{{\rm\ M}_\odot}\else{${\rm\ M}_\odot$}\fi}

\def \zsun{\ifmmode{{\rm\ Z}_\odot}\else{${\rm\ Z}_\odot$}\fi}

\newcommand{\hii}{H\,{\sc ii}{}}

\begin{document}

   \title{A meta analysis of core-collapse supernova $^{56}$Ni masses}
   \author{J.~P. Anderson\inst{1}}

   \institute{$^{1}$European Southern Observatory, Alonso de C\'ordova 3107, Casilla 19, Santiago, Chile\\
              \email{janderso@eso.org}}

\titlerunning{CC SN $^{56}$Ni mass distributions}
\authorrunning{Anderson}
   \date{}
 
  \abstract
   {A fundamental property determining the transient behaviour of core-collapse supernovae (CC~SNe) is the amount of radioactive $^{56}$Ni synthesised in the explosion. Using established methods, this is a relatively easy parameter to extract from observations.}
   {I provide a meta analysis of all published $^{56}$Ni masses for CC~SNe.}
   {Collating a total of 258 literature $^{56}$Ni masses I compare distributions of the main CC~SN types: SNe~II; SNe~IIb; SNe~Ib; SNe~Ic; and SNe~IcBL.}
   {Using these published values, I calculate a median $^{56}$Ni mass of 0.032\msun\ for SNe~II (N=115), 0.102\msun\ (N=27) for SNe~IIb, SNe~Ib = 0.163\msun\ (N=33), SNe~Ic = 0.155\msun\ (N=48), and SNe~IcBL = 0.369\msun\ (N=32). On average, stripped-enevelope SNe (SE-SNe: IIb; Ib; Ic; and Ic-BL) have much higher values than SNe~II. These observed distributions are compared to those predicted from neutrino-driven explosion models. While the SN~II distribution follows model predictions, the SE-SNe have a significant fraction of events with $^{56}$Ni masses much higher than predicted.}
   {If the majority of published $^{56}$Ni masses are to be believed, these results imply significant differences in the progenitor structures and/or explosion properties between SNe~II and SE-SNe. However, such distinct progenitor and explosion properties are not currently favoured in the literature. Alternatively, the popular methods used to estimate $^{56}$Ni masses for SE-SNe may not be accurate. Possible issues with these methods are discussed, as are the implications of true $^{56}$Ni mass differences on progenitor properties of different CC~SNe.}


   \maketitle
%

\section{Introduction}
CC~SNe are the explosive end of massive stars.
A range of physical processes power their electromagnetic output, however the vast majority of explosions show a common property: they produce a significant amount of radioactive material that powers their luminosity at some epoch.
The dominant radioactive decay chain for the first several hundred days of SN evolution is $^{56}$Ni to $^{56}$Co to $^{56}$Fe \citep{arn96}. Therefore the mass of $^{56}$Ni synthesised in the explosion is a critical parameter to understand. 
The amount of $^{56}$Ni produced in an explosion is dependent on the explosion properties and the progenitor core structure (see e.g. \citealt{ugl12,pej15c,suk16,suw19} and references therein). How the decay of this material affects the subsequent transient properties then depends on the ejecta structure, the degree to which the $^{56}$Ni is mixed outwards, and how asymmetric the $^{56}$Ni production is. Constraining the $^{56}$Ni yield in different SN types and its effect on their transient behaviour is key to further our understanding of massive star explosions.\\
\indent CC~SNe can be broadly separated into those showing strong, long-lasting hydrogen in their spectra: hydrogen-rich SNe~II, and those that do not: hydrogen-poor SE-SNe. SNe~II show a large diversity in their photometric and spectroscopic behaviour (e.g. \citealt{and14a,gut17a}). However for the purpose of this work SNe~II are not split into slower or faster decliners (IIP and IIL respectively in historical classification terminology, \citealt{bar79}), nor peculiar events. Importantly, the vast majority of SNe~II are understood to be the result of massive stars exploding with large hydrogen-rich envelopes still present, where their luminosity is powered by shock energy for the first few months post explosion. Once their initially ionised hydrogen-dominated ejected are fully recombined, their light curves show a drop in luminosity until they fall onto their radioactively powered s$_3$ decline (e.g. \citealt{and14a}).\\
\indent SE-SNe can be split into different types depending on the presence or absence of specific lines in their optical spectra (see \citealt{fil97} and \citealt{gal17} for reviews on SN classification). These differences are thought to imply a sequence of increasing envelope stripping prior to explosion. SNe~IIb do display hydrogen features in their early-time spectra, however these features disappear post maximum light and the SNe transition to be more similar to SNe~Ib. This latter class lacks hydrogen but show strong helium features. Finally, in the spectra of SNe~Ic hydrogen and helium are usually both absent \citep{mod16}\footnote{Although see the case of SN~2016coi \citep{yam17,pre18b}.}. A small subset of SNe~Ic show much broader spectral features and are thus named SNe~IcBL (broad line). While SE-SNe display diversity in their light-curve properties (absolute magnitudes, decline rates), their overall morphologies are more homogeneous than SNe~II, generally displaying a characteristic bell-shaped
luminosity evolution (see samples of light curves in e.g. \citealt{dro11,tad15,lym16,pre16,str18}). SE-SNe are therefore understood to be powered by radioactive decay \citep{arn82} for the vast majority of their evolution (the exception being the very early phases) with the decay of $^{56}$Ni powering the peak luminosity and $^{56}$Co becoming dominant several weeks later.\\
\indent The clear observational differences between SNe~II and SE-SNe imply 
significant differences in progenitor evolution to produce different pre-SN stars that then 
lead to distinct observational classifications. Initially it was thought that SE-SNe were the 
result of single Wolf-Rayet (WR) stars (more massive than 25-30\msun) where their envelopes have 
been lost through strong stellar winds (e.g. \citealt{beg86,sch87}). However, there is now mounting evidence (see e.g. \citealt{smi14c} and discussion below) that at least a significant fraction SE-SNe arise from lower mass progenitors where the envelope stripping is achieved through interaction with a close binary companion (e.g. \citealt{pod92}). In this latter scenario SE-SN progenitors have similar initial masses to those of SNe~II. Therefore, while the outer structures of progenitors at explosion epoch are different, their inner core structures should be similar. In the single massive star scenario SE-SN and SN~II progenitor cores evolve differently due to their distinct initial masses. Given that it is this core structure that determines the resulting details of explosion, constraining how these are similar or different between SN types is key for our understanding of CC~SNe.\\
\indent In the case of SNe~II, direct detections of progenitor stars on pre-explosion images has 
provided strong evidence that the majority arise from red supergiants with initial masses 
between 8 and 18\msun\ (see \citealt{sma15} and references therein). There is some suggestion of a lack of higher mass progenitors, an observation dubbed the `red supergiant problem' (\citealt{sma09}, although see e.g. \citealt{dav18}). Searches for progenitors of SE-SNe have been less conclusive. \cite{eld13} used 12 non-detections at the explosion sites of SE-SNe to argue against $\ge$25\msun\  Wolf-Rayet stars being their progenitors. However others have claimed that such progenitors would be hard to detect in the usual band-passes available, given their hot temperatures and correspondingly blue spectral energy distributions (e.g. \citealt{yoo12}).\\ 
\indent In recent years, a larger number of possible SE-SN progenitor detections have been published. \cite{van18} and \cite{kil18} studied the first possible identification of a SN~Ic progenitor (SN~2017ein). Both concluded that the visible point source was consistent with a $\ge$45\msun\ initial mass progenitor, however later observations are required to determine whether the point source is the actual progenitor or a compact cluster\footnote{Although if the source is cluster its spectral energy distribution is also consistent with the SN arising from a very massive progenitor \citep{van18}.}. In the case of SNe~Ib, there is also only one possible progenitor identification; that of iPTF13bvn. While \cite{cao13} initially argued for a massive, compact WR star progenitor, \cite{eld15} and \cite{kim15} later revised the pre-explosion photometry and concluded that the progenitor was more likely to be of lower mass (i.e. $\le$20\msun, also see \citealt{fol16}). The SN~IIb class has larger sample of progenitor detections. All such observations favour low-mass ($\le$20\msun) progenitors formed through binary interaction (SN~1993J; e.g. \citealt{mau09}, SN~2011dh; e.g. \citealt{van13}, SN~2016gkg; e.g. \citealt{tar18}). In conclusion, while the statistics are still low, the direct detection of progenitor stars on pre-SN images does not currently favour a significant difference in initial progenitor mass between SNe~II anad SE-SNe (although the only possible detection of a SN~Ic progenitor is intriguing).\\
\indent Constraints on pre-SN stars - that are then used to infer zero age main sequence (ZAMS) masses - can also be obtained through modelling SN light curves and spectral velocities. For SNe~II, while in some cases hydrodynamic modelling studies have claimed $\ge$20\msun\ progenitors (e.g. \citealt{utr17}), most published results fall within a similar mass range to those of progenitor detections\footnote{A systematic offset to higher masses from hydrodynamic modelling if often discussed in the literature (see e.g. \citealt{ber11}). While this may exist, in the main the offset is not huge and does not significantly affect the progenitor mass range for the purpose of the discussion here.} (see e.g. \citealt{bar15}). (Also note the difficulty in inferring initial progenitor mass estimates from hydrogen envelope mass/ejecta constraints, see \citealt{des19}.) SE-SN ejecta masses have been estimated using the application of both analytical and hydrodynamic modelling, with results consistently arguing for low mass ejecta and therefore low initial progenitor masses consistent with binary evolution and inconsistent with most single-star evolution models (see \citealt{dro11,tad15,lym16,tad18b,ber18,pre19}). However, there are suggestions that the light curves of some SNe~Ic constrain their progenitors to be of higher initial mass (see e.g. \citealt{val12,tad16}).\\
\indent Less direct constraints on SN progenitors come from analyses of the environments in which they are discovered. Studies of resolved stellar populations surrounding SN explosion sites \citep{mau17,mau18} have argued for a decreasing progenitor age sequence from SNe~II through SNe~IIb, Ib, and Ic, i.e. an increasing progenitor mass sequence: SNe~II-IIb-Ib-Ic. Further afield, studies of unresolved stellar populations have also suggested parent stellar population age differences, with SNe~Ic consistently being found within younger stellar populations, and therefore arising from higher mass progenitors then other CC~SN types \citep{and12,kan17,kun18,gal18}. Finally, SNe~IcBL and specifically those accompanying long-duration gamma ray bursts (GRBs), have environments consistent with coming form the highest mass progenitors of all SNe discussed here \citep{kel08,kel12}.\\
\indent In addition to constraints on progenitor ages, environment studies have investigated parent stellar population oxygen abundances, which are then used as 
progenitor metallicity proxies (\citealt{mod08,and10,lel11,san12,kun18,gal18,mod19}). Most CC~SN types do not show significant differences, however SNe~Ic are consistently found in higher metallicity regions than SNe~II and SNe~Ib, while SNe~IcBL (see \citealt{mod19}) are systemtically found in galaxies of lower metal abundance than SNe~Ic and all other CC~SNe. Such metallicity differences may be key in explaining the origin of some of the diversity within the CC~SN family.
These environment studies (both resolved and unresolved) are somewhat in contradiction to the statistical studies of the ejecta masses of SE-SNe discussed above: environment studies suggest significant mass differences between some CC~SN types, while estimated ejecta masses suggest very similar initial progenitor masses between SNe~II and SE-SNe.\\
\indent Progenitor mass constraints can also be derived from spectroscopy of SNe at nebular times (several hundred days post explosion) when the ejecta has become optically thin and the inner core material is revealed. Observations at these epochs constrain the core mass that is predicted to strongly correlate with the ZAMS mass \citep{woo02}. Nebular spectroscopic studies have generally concluded for SN~II progenitor masses similar to those from direct detections (see e.g. \citealt{jer15,val16}, although see \citealt{and18a}). Nebular-phase observations of SE-SNe are less common, but the statistical analysis of \cite{fan18} concluded that SNe~Ic appear to show spectra consistent with more massive progenitors than SNe~Ib.\\
\indent SNe~IcBL and those associated with GRBs (in particular) show specific properties that may differentiate them from other CC~SNe. Specifically, (as above) the broad-line designation means large ejecta velocities and this implies significantly higher energy explosions than other CC~SN types. Significant asphericities have also been discussed to explain the properties of SNe~IcBL \citep{mae03,des17}. Such high explosion energies are probably inconsistent with these events arising from standard explosion mechanisms (see further discussion below), which brings caveats to discussing their analysis in a similar vain to other types.\\
\indent In summary, there is a large body of work attempting to constrain CC~SN progenitor mass differences. 
Most studies suggest significant overlap in the ZAMS mass of SNe~II and SE-SNe. This suggests that at least a significant fraction of SE-SNe arise through binary evolution (stripping the outer envelope) but from stars with initially very similar masses to SNe~II. At the same time, observations do suggest that some SNe~Ic arise from more massive progenitors that may evolve as single stars (i.e. they may be massive enough to lose their envelope through stellar winds). Further constraints are required to shed light on progenitor differences between CC~SN types and to constrain the explosion mechanism.\\
\\
\indent As outlined earlier, a critical parameter for understanding how a SN explosion proceeds is the mass of synthesised $^{56}$Ni. \cite{mul17} compared a sample of observed SN~II $^{56}$Ni masses with those from explosion models \citep{suk16} finding consistent results.
Additional observational samples of SN~II $^{56}$Ni masses can be found in e.g.:
\cite{ham03a,and14a,spi14,val16}. SE-SNe are generally analysed separately from SNe~II, and various work has provided distributions of $^{56}$Ni masses, see e.g. \cite{dro11,lym16,pre16,tad15,tad18}. These latter studies present distributions with higher mean $^{56}$Ni values than those of SNe~II. However, there does not appear to be any formal study analysing SN~II and SE-SN $^{56}$Ni mass distributions in comparison. This is the aim of the current paper: to provide a meta analysis of CC~SN $^{56}$Ni masses to investigate whether there exist statistical differences between different CC~SN types and discuss the implications that follow.\\
\indent This paper is organised as follows. In the next section I outline how the meta sample is produced and summairse its characteristics. Section 3 discusses the main methods used in the literature for estimating $^{56}$Ni masses. In Section 4 I present $^{56}$Ni masses for: SN~II; SN~IIb, SN~Ib; SN~Ic; SN~IcBL (including GRB-SNe), and compare these statistically. This is followed by a discussion of possible explanations of these results including caveats on the commonly used methods for $^{56}$Ni mass estimation in Section 5. Final conclusions are presented in Section 6.\\

\section{A meta sample of CC~SN $^{56}$Ni masses}
The aim of this work is to collate and then compare literature CC~SN $^{56}$Ni masses. In the later discussion section I outline possible systematics in how these values are estimated and how these may affect my results and conclusions. However, here I do not attempt to remeasure $^{56}$Ni masses. Following this there is no preference for any specific method for estimating these values. To produce this sample I simply searched the literature for all published masses. Specifically, the SAO/NASA ADS astronomy query form\footnote{\href{http://adsabs.harvard.edu/abstract_service.html}{http://adsabs.harvard.edu/abstract\_service.html}} was used, searching for articles with `supernova' and `type II', then `supernova' and `type IIb' and so forth in manuscript abstracts. The resulting lists of articles were then scanned to find published $^{56}$Ni masses. While the vast majority of values have probably been collated (through to August 2018)\footnote{Two additional samples were published after this date. \cite{pre19} published a sample of SE~SNe, while \cite{tad18b} published a sample of SNe~IcBL. These additional values are in line with those presented for the meta sample here, and their inclusion would not affect the results nor conclusions of the present study.}, it is probable that some have been missed. In the many cases of multiple
literature values for individual SNe a mean mass is calculated. A full reference list for all $^{56}$Ni masses can be found in the appendix. 
The two main methods for estimating $^{56}$Ni in SNe~II and SE-SNe are outlined next.\\

\section{Standard procedures to estimate $^{56}$Ni masses}
While there are many variations on the exact details of how authors estimate $^{56}$Ni masses in SNe, there are two standard methods. First, in SNe~II the tail luminosity is used as a robust estimate of the synthesised mass of radioactive material. Once the hydrogen ejecta has fully recombined the light curves of SNe~II fall down to their s$_3$ radioactive tails. At this stage their luminosity is powered by the decay of $^{56}$Co. This is well established (e.g. \citealt{woo88}) and the s$_3$ slope follows the predicted decline (from the rate of decay of $^{56}$Co) extremely well in most SNe~II (as the radioactive emission is fully trapped by the ejecta, see \citealt{and14a} for a discussion of SNe~II where this is not the case). It is therefore reasonably straightforward to extract the $^{56}$Ni mass using the bolometric luminosity of a SN~II if one also has reasonable constraints on the explosion epoch (see e.g. \citealt{ham03a} for an outline of one possible procedure).\\
\indent In the case of SE~SNe the tail in the light curves generally decline significantly quicker than that predicted by the $^{56}$Co decay rate \citep{whe15} due to incomplete trapping\footnote{Although 20-30 days post maxmium most of the emission is trapped and it may be possible to use this epoch in the same way as for SNe~II to estimate $^{56}$Ni masses \citep{des15b}.}. Therefore the tail luminosity has not been used for SE-SN $^{56}$Ni estimates. SE-SN $^{56}$Ni masses are generally estimated using `Arnett's rule' \citep{arn82}\footnote{For a fraction of the current meta sample SE-SN $^{56}$Ni masses are estimated through spectral modelling of nebular-phase observations (see e.g. \citealt{maz09}), or through modeling of observations around maximum light (see e.g. \citealt{ber18}).}. Arnett's rule states that the SN luminosity at peak brightness equals the instantaneous rate of energy deposition (from radioactive decay). This assumes that radioactive decay is the only energy source at peak magnitude and
involves a number of assumptions that are outlined and questioned in \cite{kha18} and \cite{des16} (as will be discussed later). The method used for SE-SNe is therefore more indirect than that for SNe~II.\\
\indent It is important to note the various uncertainties involved in $^{56}$Ni mass estimations. The above methods require that observed photometry are converted into bolometric luminosities. Measurements have to be corrected for line of sight extinction, distance, and the missing flux outiside the observed wavebands. All three of these corrections can have significant uncertainties. In addition, photometry may be contaminated by underlying galaxy light. These observational errors then sum with those arising from the $^{56}$Ni estimation method. The latter assume an exclusive contribution from $^{56}$Ni together with spherical asymmetry. As noted above, Arnett's rule also has a range of assumptions that are not necessarily valid for all SNe.\\
\indent How these uncertainties may affect the results presented in this work is further discussed in Section~\ref{discussion}. However, it appears unlikely that they can explain all the differences in $^{56}$Ni masses between CC~SNe that will now be presented.

\begin{figure*}
\begin{center}
\includegraphics[width=15cm]{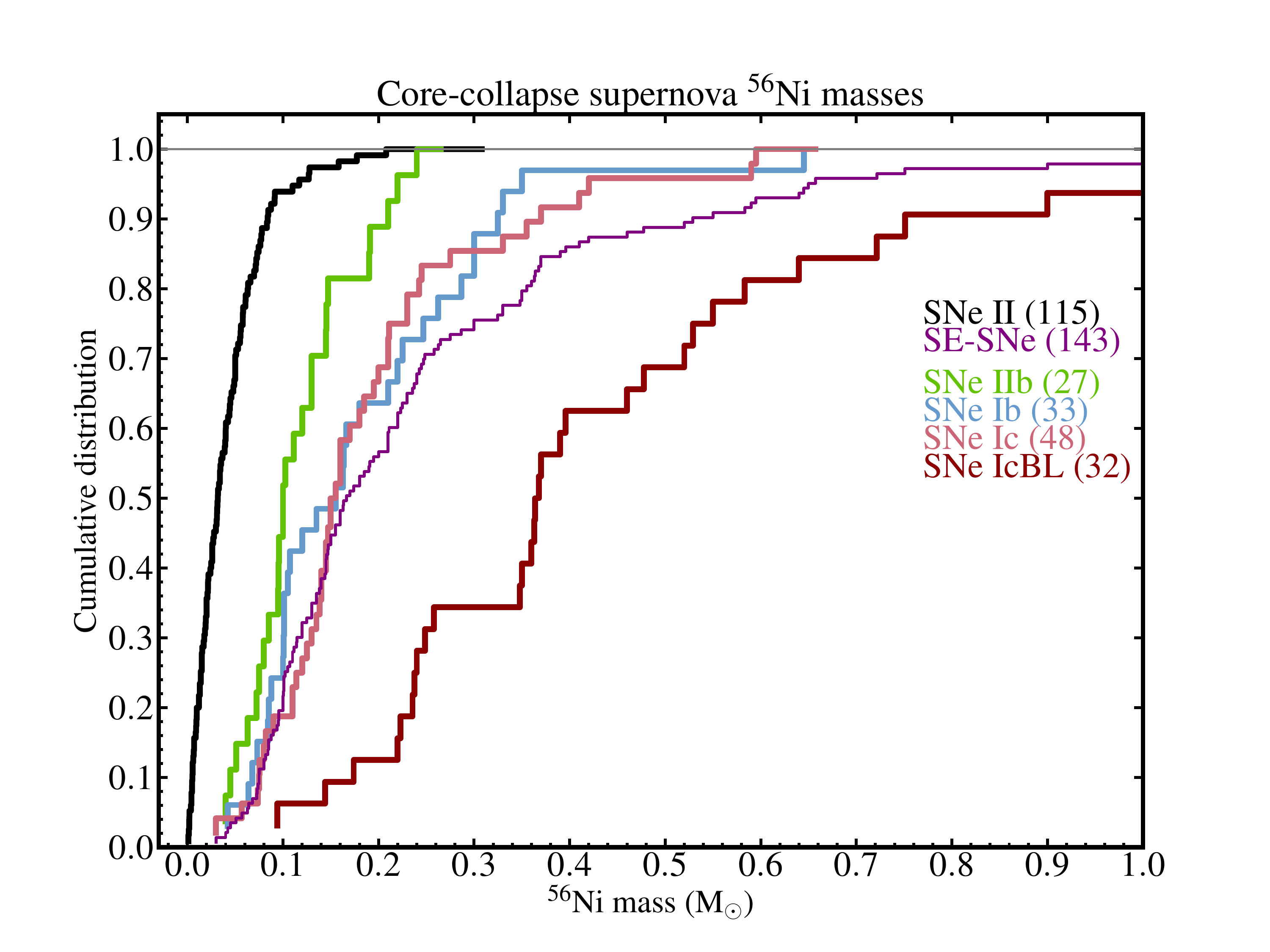}
\caption{Cumulative distributions of the $^{56}$Ni masses for the different CC~SN types analysed in this study. (The SNe~IcBL - that includes GRB-SNe - do not complete their cumulative distribution in this figure given that a small number of events are estimated to have synthesised more than 1\msun\ of $^{56}$Ni.)}
\label{ni56plotall}
\end{center}
\end{figure*}

\begin{table*}
\begin{center}
\begin{tabular}[t]{cccccc}
\hline
SN distribution (N) & Mean (\!\msun) & Standard deviation (\!\msun) & Median (\!\msun) & Max (\!\msun) & Min (\!\msun)\\
\hline	
SN~II (115) & 0.044 & 0.044 & 0.032 & 0.360 & 0.001\\
SE-SN (143) & 0.293 & 0.295 & 0.184 & 2.400 & 0.030\\
\hline
SN~IIb (27) & 0.124 & 0.061 & 0.102 &  0.280 & 0.030\\
SN~Ib (33) &  0.199 & 0.146 & 0.163 & 0.920 & 0.030\\
SN~Ic (48) & 0.198 & 0.139 & 0.155 & 0.840 & 0.030\\
SN~IcBL (32) & 0.507 & 0.410 & 0.369 & 2.400 & 0.070\\
\hline        
\hline
\end{tabular}  
\caption{CC~SN $^{56}$Ni statistics. In the first column I list the SN distribution and the number of events within that distribution in brackets. Means, standard deviations, and medians of the distributions are then presented in the second, third and fourth columns respectively. In the last two columns I list the minimum and maximum value in each distribution (these latter values are those from the individual literature measurements, while in Fig.~\ref{ni56plotall} the mean values for each SN are plotted).}
\label{tabval}
\end{center}
\end{table*}

\begin{table}
\begin{center}
\begin{tabular}[t]{cccc}
\hline
SN distributions (N) & D & p\\
\hline	
SN~II (115)--SE-SNe (143) & 0.773 & 1.2$\times$10$^{-34}$\\
\hline
SN~II (115)--SN~IIb (27)  & 0.687 & 6.1$\times$10$^{-10}$\\
SN~II (115)--SN~Ib (33)   & 0.774 & 1.5$\times$10$^{-14}$\\
SN~II (115)--SN~Ic (48)   & 0.801 & 3.0$\times$10$^{-20}$\\
\hline
SN~IIb (27)--SN~Ib (33)   & 0.323 & 0.07\\
SN~IIb (27)--SN~Ic (48)   & 0.375 & 0.01\\
SN~Ib (33)--SN~Ic (48)    & 0.227 & 0.23\\
SN~Ic (48)--SN~IcBL (32)  & 0.635 & 1.3$\times$10$^{-7}$\\
\hline              
\hline
\end{tabular}  
\caption{KS statistical differences between the various different CC~SN $^{56}$Ni distributions. In the first column the two distributions being compared are listed together with the number of events within each distribution in brackets. In the third column the D parameter is given while in the last column the p value is presented.}
\label{tabks}
\end{center}
\end{table}

\section{Results}
Figure~\ref{ni56plotall} presents cumulative $^{56}$Ni distributions of all CC~SNe. Mean, median and standard deviations for each are presented in Table~\ref{tabval}. Table~\ref{tabks} lists the Kolmogorov-Smirnov (KS) statistics between the $^{56}$Ni mass distributions of the different types. The SN~II distribution is clearly skewed towards significantly lower $^{56}$Ni masses than SE-SNe. \textit{The median value for the SNe~II is a factor of three lower than the SNe~IIb and a factor of five lower than the SNe~Ib and SNe~Ic}. The SN~II distribution is statistically distinct from all other distributions (Table~\ref{tabks}).\\ 
\indent Inspecting the minimum and maximum values for each class is also particularly telling: the mininum $^{56}$Ni mass for any SE-SN is 0.03\msun, while 45\%\ of SNe~II have lower values, the lowest being 0.001\msun \footnote{This may suggest an observational bias against detecting SE-SNe with low $^{56}$Ni masses. Such SNe would be much dimmer than the observed SE-SN population and may have gone undetected. If this is the case, then we should expect the current and future generation of discovery surveys - that search deeper and with a higher cadence than ever before - to start detecting such events.}. At the higher end of the SN~II distribution there is some overlap with SE-SNe: 100\%\ of the SNe~IIb overlap with the SNe~II; 76\%\ of the SNe~Ib; and 60\%\ of the SNe~Ic. In general, it is the lack of low SE-SN $^{56}$Ni values that produces such statistically different distributions. The mean value for the $^{56}$Ni mass of SN~1987A (the nearest SN to Earth in the last $\sim$400 years) is 0.072\msun\ and this is often used as the canonical value for SNe~II. However, this value is a factor of two higher than the median $^{56}$Ni mass of SNe~II in the current sample. Indeed, 83\%\ of the SNe~II have estimated values lower than SN~1987A. All six events classified as `87A-like' have relatively high $^{56}$Ni estimates, with a mean value of 0.086\msun.\\
\indent With respect to differences between the SE-SN classes, the SNe~IIb appear to produce less $^{56}$Ni than the other types, while there is no statistically significant difference between the SNe~Ib and SNe~Ic. Meanwhile, the SN~IcBL distribution (where GRB-SNe are also included) contains by far the largest number of high masses, with many SNe~IcBL estimated to have synthesised more than 0.5\msun. To put this in context, Type Ia SNe produce on average 0.6\msun\ (e.g. \citealt{sca19}). As shown in Table~\ref{tabval} and Fig.~\ref{ni56plotall} the SNe~Ib and SNe~Ic distributions also contain such high values. To see how much these drive the differences seen between SNe~II and SE-SNe, the average $^{56}$Ni values are recalculated after removing $>$\,0.35\msun\ values. The SNe~Ib and SNe~Ic mean $^{56}$Ni are still a factor of four larger than the SNe~II.

\subsection{Comparison to model predictions from neutrino driven explosions}
The most commonly studied and favoured explosion mechanism for CC~SNe is the
so-called `neutrino-driven explosion' (see e.g. \citealt{jan17} for a recent review). Core-collapse is initiated when the iron core becomes too massive to support itself (through degeneracy pressure) against gravity. When the material from this resulting collapse reaches nuclear densities the collapse is halted and a shock wave drives through the still inwardly falling outer layers of the core. It has long been accepted that the initial shock from the core bounce stagnates. In the neutrino driven mechanism, the huge neutrino flux from the accreting proto-neutron star - assisted by turbulent motions - revives the shock and produces the CC~SNe we observe. However, while some simulations have succeeded in producing successful explosions (e.g. \citealt{mul12}), many have not (see discussion in e.g. \citealt{tak14b}), and even those that do produce low-energy events. For these reasons, others have proposed alternative explosion mechanisms (see e.g. \citealt{pap15}). A full exploration of alternative models in the context of $^{56}$Ni production is beyond the scope of this paper, but the reader should be aware that the below comparisons are made with only a specific theoretical framework that may not be the dominant one in Nature.\\
\indent A number of studies have provided $^{56}$Ni yields from neutrino-driven explosion models of various levels of complexity (\citealt{ugl12,pej15c,suk16,suw19}\footnote{Most of these works only model the initial explosion and the estimated nucleosynthesis; they do not go on to produce light curves nor spectra that can be compared to observed events.
\cite{suk16} do produce such observables, and their explosions struggle to reproduce observed SE-SNe.}, $^{56}$Ni yields from other models are discussed below).
While the exact results vary between studies (e.g. the dependence of the synthesised $^{56}$Ni on explosion energy and initial progenitor mass), the 
predicted $^{56}$Ni mass range is somewhat similar, as is the
maximum predicted value. In the case of SNe~II, \cite{mul17} showed that the $^{56}$Ni distribution 
predicted by \cite{suk16} was in reasonable agreement with observed SNe~II, and qualitatively we 
find the same result here with a much larger sample. SE-SNe however, have estimated values in significant 
excess of model predictions. The maximum predicted value in each of the four explosion-model studies cited 
above is: 0.15\msun\ \citep{ugl12}; 0.2\msun\ \citep{pej15c}; 0.171\msun\ \citep{suk16}; and 0.226\msun\ \citep{suw19}. 
Taking the most extreme of these (0.226\msun), 30\%\ of SN~Ib have higher inferred values, 
together with 29\%\ of SNe~Ic and 84\%\ of SNe~IcBL (only 3\%\ of SN~II and 7\%\ of SNe~IIb 
have values in excess of this limit). Therefore, even in the most optimistic case (that many of the SE-SNe 
come from the highest values from explosion model predictions), a significant fraction of $^{56}$Ni estimated 
values for SE-SNe are outside of neutrino-driven model estimations\footnote{Although see \cite{ume08} for 
much higher $^{56}$Ni masses produced from SNe more massive than 30\msun.}. As Fig.~\ref{ni56plotall} 
shows, a number of SE-SNe have values twice or even three times are large as this limit. 
The implications of these results, together with a series of caveats and possible explanations, are now discussed.

\section{Discussion}
\label{discussion}
Two main results are presented in this paper: 1) that values of $^{56}$Ni in the literature are systematically larger for SE-SNe than for SNe~II, and 2) that a significant fraction of SE-SNe have published $^{56}$Ni masses in excess of the largest masses predicted by a range of different neutrino-driven explosion models. 1) implies significant differences in the progenitor structures and explosion properties between hydrogen-rich and hydrogen-poor CC~SNe. However, 2) may suggest that the estimates for many SE-SNe are significantly in error, or that the currently most popular explosion model (neutrino-driven explosion) is not applicable to a significant fraction of CC~SNe.\\
\indent The general current consensus (although it is still debated) is that a significant fraction - if not the vast majority - of SE-SNe arise from binary systems where mass transfer is responsible for removing the outer hydrogen (type IIb, Ib) and helium (Ic) rich layers of the progenitors. In this hypothesis, the initial progenitor masses of SNe~II and SE-SNe are similar (although those of SE-SNe are still probably higher on average). While their pre-SN outer structures are distinct, their core-structures should be somewhat indistinguishable (in most cases the core will evolve independently of surface processes). Therefore it is not clear how one could arrive at such distinct $^{56}$Ni masses (as presented in this study) if the progenitors have similar initial masses. If one postulates that SE-SNe actually have significantly more massive progenitors (and evolve either as single stars or in binary systems) then one may speculate that their cores have more material at sufficiently high densities to produce higher amounts of $^{56}$Ni during the explosion. However, this is not actually predicted by neutrino-driven explosion models \citep{ugl12,pej15c,suk16,suw19}: a 25\msun\ star does not necessarily produce more $^{56}$Ni than an e.g. 15\msun\ star, and models do not produce $^{56}$Ni masses in excess of 0.2\msun. In summary, \textit{if literature $^{56}$Ni values are to be believed, then a) the progenitor structures (and by inference the initial progenitor properties) must be significantly more different between SNe~II and SE-SNe that currently believed, and b) progenitor structures and/or explosion properties of SE-SNe must be distinct from those predicted by stellar evolution and currently favoured explosion models}.\\ 
\indent There are various sources of observational error that may affect $^{56}$Ni mass estimates: 1) errors in photometry; 2) errors in SN distances; 3) errors in extinction corrections; and 4) errors in bolometric corrections. If any of these corrections are systematically wrong for one SN type compared to another, this may produce some of the differences we observe. The errors from 1) can be assumed to be negligible compared to the rest. There is no reason to believe that errors in distances (2) are any different between SN types (although very large or very small individual $^{56}$Ni masses may be due to distance errors).\\ 
\indent Estimation of accurate host galaxy extinction for CC~SNe is notoriously difficult (see recent discussion for the case of SNe~II: \citealt{dej18}). Such estimates are generally derived from assuming some uniform intrinsic colour for a given SN type and correcting assumed reddened SNe to assumed unreddened events (see e.g. \citealt{str18}), or by measuring the strength of absorption from interstellar sodium (see e.g. \citealt{phi13}). However, both of these methods have strong caveats.
To investigate how these errors may affect the results from the current study, I also compile host galaxy extinction estimates employed by each study contributing $^{56}$Ni masses to this work. 
The median host galaxy visual extinction does increase from hydrogen-rich through the stripped-envelope classes: 0.12\,mag for SNe~II; 0.25\,mag SNe~IIb; 0.30\,mag SNe~Ib; 0.56\,mag SNe~Ic, while for the SNe~IcBL the median is 0.12\,mag. However, this is somewhat expected, given the increasing sequence of association from SNe~II through to SNe~Ic to bright \hii\ regions (\citealt{and12}, where one may expect higher levels of extinction). Still, in the extreme case of assuming that the host extinction estimates are systematically wrong between e.g. SNe~II and SNe~Ic, one can estimate that the addition of 0.44 visual magnitudes (the difference in the median values of SNe~II and SNe~Ic) to the bolometric luminosity of a SN~II would only increase the median $^{56}$Ni value to 0.048\msun. In order to increase this to the median of the SNe~Ic (0.155\msun), a difference of 1.7 mag is required. It seems extremely unlikely that such a difference has gone unnoticed.\\ 
\indent Finally, if there were a systematic error in the way bolometric corrections were applied between SNe~II and SE-SNe this could produce some difference in $^{56}$Ni masses. Bolometric corrections are either produced by using very well observed (in wavelength) SNe as templates and assuming similar spectral energy distributions for less well observed events, or through comparison to models. In this study $^{56}$Ni masses are compiled from many different sources and in many SNe there are multiple (up to eight) values from different studies. Comparing these values can give some idea of the differences resulting from different bolometric corrections. In the case of SNe~II the mean standard deviation of multiple $^{56}$Ni mass estimates is 0.013\msun, while for e.g. SNe~Ic it is 0.06\msun. If one assumes that this larger scatter for the SNe~Ic is due to systematically incorrect bolometric corrections for some of the estimates for these SNe, then one can speculate that such an error propogates to differences in $^{56}$Ni mass estimates. However, the above difference is not sufficient to explain the significantly distinct $^{56}$Ni distributions. In summary, there does not appear to be any significantly large observational error that would negate the conclusion of large differences between the estimated $^{56}$Ni masses of SNe~II and SE-SNe.\\

\subsection{Alternative explanations}
The methodology for estimating $^{56}$Ni masses from SN~II observations is understood to be robust in a theoretical sense. However, doubt has often been cast as to the accuracy of using Arnett's rule in the case of SE-SNe (the methodology used for the vast majority of literature values compiled here). 
\cite{kat13} published an `exact integral relation between the $^{56}$Ni mass and the bolometric light curve', arguing that this overcomes issues with Arnett's rule such as assumed opacities, density distribution, and the $^{56}$Ni deposition distribution into the ejecta.
\cite{des15b,des16} published radiative transfer models of SE-SNe (IIb/Ib/Ic) 
based on explosions of the mass donor in a close-binary system. These authors 
concluded that for this set of models Arnett's rule overestimates the $^{56}$Ni
mass by around 50\%\ (while suggesting that the \citealt{kat13} procedure yields
more reliable values). However, these models (and explosion models of SE-SNe in general, see e.g. \citealt{suk16} for another example) often struggle to accurately reproduce observations, with e.g. the model rise times being longer than those observed.
More recently, \cite{kha18} discussed in detail the assumptions that are contained within the classical Arnett model and argued - using comparison to numerical simulations - that Arnett's rule does not hold in general (while presenting new analytic relations), only being valid in certain explosion configurations. It is important to stress: the difference in published values between SNe~II and SNe~Ib/SNe~Ic is a factor of five (Section 4). This is significantly larger than the offsets discussed in e.g. \cite{des15b} and \cite{kha18}.\\
\indent As discussed above, the neutrino-driven explosion mechanism may not be at play in all CC~SNe. \cite{sok18} argues that jets are required to explode many if not all CC~SNe, and that a paradigm shift is needed to move away from the neutrino-driven mechanism to a `jet-driven explosion mechanism that is aided by neutrino heating'. However, there are few studies of nucleosynthetic yields from non-standard explosion models (to my knowledge). \cite{bar18} presented a model of a GRB central engine producing the accompanying SN~IcBL through a jet-driven aspherical explosion. (Aspherical explosions may be applicable to the majority of SE-SNe, given the number of such events showing double-peaked nebular-phase spectral features, \citealt{mae08}.) This model calculated the synthesised $^{56}$Ni mass through a temperature condition, and estimated a yield of 0.24\msun\ of $^{56}$Ni. This is higher than for any of the explosion models discussed above, but it should be noted that there are many SE-SNe (and not just the IcBL) in Fig.\ref{ni56plotall} that have values in excess of this. Additional explosion and nucleosyhthesis modelling is required to understand whether alternative models can indeed consistently produce $^{56}$Ni masses much higher than 0.2\msun.\\
\indent In the case of SNe~IcBL estimated $^{56}$Ni masses are extremely large, being of the same order as those for SNe~Ia, together with a tail out to values higher than 1\msun. This seems to be beyond any available progenitor exploded through the neutrino-driven explosion model. As noted previously, one possibility is that SN~IcBL explosions are significantly asymmetric, and we observe these events as SNe~IcBL when the explosion direction is towards the observer (see e.g. \citealt{mae03,des17b}). In such cases $^{56}$Ni
masses derived assuming spherical symmetry (e.g. Arnett's rule) will be overestimated (\citealt{des17b}, and ejecta masses will be underestimated). This explanation cannot work for the SE-SN sample at large as statistically one would expect to observe both high and low values depending on the viewing angle of the observer.\\
\indent Finally, a remaining explanation is simply that for many SE-SNe radioactive decay is not the dominant power source of their luminosity. If an additional power source is present (e.g. a magnetar) then assuming that all the power comes exclusively from $^{56}$Ni will lead to an overestimate of such masses.

\section{Conclusions}
Using published $^{56}$Ni masses from the literature that employ standard methods for estimating the contribution of radioactive decay power to SN luminosities, I have shown that hydrogen-poor SE-SNe are estimated to produce around five times more $^{56}$Ni than hydrogen-rich SNe~II. This difference is highly statistically significant. While the distribution of $^{56}$Ni for SNe~II agrees with predictions from
neutrino-driven explosion models, that of SE-SNe does not.
These results imply that either SE-SN progenitors and their subsequent explosion are significantly distinct from those of SNe~II, or that there is a systematic error in how $^{56}$Ni masses are calculated. This work serves to highlight these issues to the community. The amount of radioactive material synthesised in CC~SNe is a critical parameter in understanding these explosive events. Therefore a detailed understanding of how well we derive such a property is vital to move our understanding forward.

\begin{acknowledgements}
The anonymous referee is thanked for their comments and suggestions.
Luc Dessart is thanked for providing detailed comments on a draft version of this manuscript. In addition, I acknowledge fruitful conversations with the following people that aided in the discussion sections of this paper: Takashi Moriya; Ond\v rej Pejcha; Tomas M\" uller; Jos\' e Luis Prieto, Keiichi Maeda; Melina Bersten, Peter Hoeflich.  
\end{acknowledgements}

\bibliographystyle{aa}


\appendix{}
\section{Reference list for $^{56}$Ni mass values}
Here all references used to compile the meta sample of CC~SN $^{56}$Ni values used in this work are listed.\\
\indent SNe~II: \cite{arn89,tur98,nad03,ham03a,elm03,zam03,pas04,hen05,pas05,vin06,gur08,pas09,kle11,ber11,and11_2,roy11a,roy11b,tad12,pas12,ost12,ins13a,bos13,spi14,pej15a,hua15,bos15,jer15,bar15,tak15,val15,yua16,dhu16,val16,ter16,gut17a,mul17,lis17,tom18,das18,and18a,hua18,hos18,tar18,tsv18}.\\
\indent SE-SNe: \cite{shi94,iwa94,iwa98,clo00,nak01,str02,fol03b,maz03,maz04,anu05,den05,maz06,ric06,tom06,tau06,fol06_2,sau06,tom08,val08,maz08,sil09,maz09,sah09,ham09,rom09,str09,dro11,elm11,tau11,ben11,pig11,val11,ber11_2,cor11,oat12,ber12,san12b,val12,buf12,ben12,mel12,mil13,tak13,mil13b,kum13,roy13,buf14,sri14,fol14_2a,wal14,mel14,che14_2,mor15,tad15,del15,liu15,mil15,pre16,roy16,sza16,fre16,lym16,pia17,maz17,cor17,can17,tad18,gan18,ber18,zha18,kum18,sah18}\footnote{From \cite{pre16} the `fully bolometric' values are used.}

\end{document}